\documentclass[a4paper,10pt,twoside]{cpc-hepnp}

\usepackage{multicol}
\usepackage{graphicx}
\usepackage{booktabs}
\usepackage{amssymb,bm,mathrsfs,bbm,amscd}
\usepackage[tbtags]{amsmath}
\usepackage{lastpage}
\usepackage{CJK}
\usepackage{floatrow}
\begin{document}
\begin{CJK*}{GB}{gbsn}

\fancyhead[c]{\small Chinese Physics C~~~Vol. 42, No. 5 (2018)
054101} \fancyfoot[C]{\small 054101-\thepage}

\footnotetext[0]{Received 6 February 2018, Published online 20 April 2018$^*$}

\title{
Calculation of multidimensional potential energy surfaces for          even-even
      transuranium nuclei: systematic investigation \\ of the triaxiality effect on the fission barrier
\thanks{Supported by  the National Natural Science Foundation of China (Nos.
 11675148 and 11505157), the Project of Youth Backbone Teachers
  of Colleges and Universities of Henan Province (No. 2017GGJS008), the Foundation and
 Advanced Technology Research Program of Henan Province (No. 162300410222),
 the Outstanding Young Talent Research Fund of Zhengzhou University (No. 1521317002) and
the Physics Research and Development Program of Zhengzhou University (No. 32410017).
}
}

\author{%
      Qing-Zhen Chai (²ñÇåìõ)$^{1}$%
\quad Wei-Juan Zhao (ÕÔά¾ê)$^{1}$
\quad Min-Liang Liu (ÁøÃôÁ¼)$^{2}$
\quad Hua-Lei Wang (Íõ»ªÀÚ)$^{1;1)}$\email{wanghualei@zzu.edu.cn}
}
\maketitle

%%%%%%%%%%%%%%%%%%%%%%%%%%%%%%%%%%%%%%%%%%%%%%%%%%%%%%%%%%%%%%%%%%%%%%%%%%%%%%%%
\address{%
$^1$
School of Physics and Engineering, Zhengzhou University,
Zhengzhou 450001, China \\
$^2$
Institute of Modern Physics, Chinese Academy of Sciences,
Lanzhou 730000, China \\
}

%%%%%%%%%%%%%%%%%%%%%%%%%%%%%%%%%%%%%%%%%%%%%%%%%%%%%%%%%%%%%%%%%%%%%%%%%%%%%%%%
\begin{abstract}
Static fission barriers for 95 even-even transuranium nuclei with charge  number           $Z=94-118$ have been systematically investigated by means of pairing  self-consistent Woods-Saxon-Strutinsky calculations using the potential energy          surface approach in multidimensional ($\beta_2$, $\gamma$, $\beta_4$)    deformation space.
Taking the heavier $^{252}$Cf nucleus (with the available fission barrier         from        experiment) as an example, the formation of the fission barrier and the
       influence of macroscopic, shell and pairing correction energies on it are                                                            analyzed.
The results of the present calculated $\beta_2$ values and barrier heights       are
                   compared with previous calculations and available experiments.
The role of triaxiality in the region of the first saddle is           discussed.
It is found that the second fission barrier is also considerably affected     by
the  triaxial deformation degree of freedom in some nuclei (e.g., the   $Z=112-118$
                                                                       isotopes).
Based on the potential energy curves, general trends of the evolution of     the
     fission barrier heights and widths as a function of the nucleon numbers are
                                                                    investigated.
In addition, the effects of Woods-Saxon potential parameter modifications (e.g.,
    the strength of the spin-orbit coupling and the nuclear surface diffuseness)
                                        on the fission barrier are briefly discussed.
\end{abstract}


\begin{pacs}
 21.10.Re, 21.60.Cs, 21.60.Ev
 \quad \quad 
$\textbf{DOI}$: 10.1088/1674-1137/42/5/054101
\end{pacs}

\footnotetext[0]{\hspace*{-3mm}\raisebox{0.3ex}{$\scriptstyle\copyright$}2018
Chinese Physical Society and the Institute of High Energy Physics of
the Chinese Academy of Sciences and the Institute
of Modern Physics of the Chinese Academy of Sciences and IOP Publishing Ltd}%

\begin{multicols}{2}

%%%%%%%%%%%%%%%%%%%%%%%%%%%%%%%%%%%%%%%%%%%%%%%%%%%%%%%%%%%%%%%%%%%%%%%%%%%%%%%%
\section{Introduction}\label{sec.I}

Transuranium nuclei are produced artificially in heavy-ion induced       nuclear
                                                                fusion reactions.
The production rates by present production paths are allowed on                an
       atom-at-a-minute to atom-at-a-month scale applying currently experimental
                                                                      techniques.
There is no doubt, with the development of radioactive beam facilities,
      heavy-ion accelerators, and highly effective detector arrays, that there will be
    increasing interest in further attacking transuranium nuclei, especially the
                  island of stability of superheavy elements~\cite{Andreyev2018}.
Both physics and chemistry have arrived on the shore of this fascinating  island
                    in experiments~\cite{Eichler2013,Hofmann2000,Oganessian2017}.
As one of the important decay modes of a heavy or superheavy nucleus,         the
spontaneous fission channel is dominated essentially by the barrier size
                                                  and shape~\cite{Hebberger2017}.
Also, the formation probability of such a nucleus in    heavy-ion-fusion
    reactions is directly related to the fission barrier, generally including its
   height and full-width at half-maximum (FWHM), since the barrier is a decisive
        quantity in the competition between neutron evaporation and fission of a
                              compound nucleus during the process of its cooling.
The production cross section of such fissioning nuclei has a           sensitive
     dependence on the barrier shape and height. For instance, a 1~MeV change in
     the fission barrier may result in a difference of several orders of magnitude in  survival probability.
Moreover, the fission barrier of very neutron-rich nuclei can affect         the
           $r$-process of stellar nucleosynthesis~\cite{Arnould1999,Mamdouh2001}.
Therefore, the fission barrier is a critical quantity for understanding      the
                                                       questions mentioned above.
%1234567890123456789012345678901234567890123456789012345678901234567890123456789

Accurately describing fission been a long standing problem ever since it was deduced, for the first time, by barrier penetration about 80 years ago~\cite{Bohr1939}. More recently, considerable effort has been made to understand the fission  problem
                                                   in both theory and experiment.
Some empirical barriers in transuranium nuclei have been              determined
  experimentally~\cite{RIPL2009} and various theoretical approaches have been used for the study of the fission barriers.
It has been found that different deformation degrees of freedom can have    different
        influence on the inner and outer barriers, though the impact may strongly
               depend on the proton and neutron numbers and the employed models.
For instance, the heights of the inner and outer barriers can usually be lowered
                                    by the triaxiality and octupole correlation,
                                   respectively~\cite{Pashkevich1969,Moller1970}.
Further, it was recently found that the outer barriers can also be lowered    by
                                 the triaxiality compared with axially symmetric
                         results in the actinide region~\cite{LuBN2012,LuBN2014}.
The most fundamental way to determine the nuclear properties is to start with  a
  real nucleon-nucleon interaction and solve the appropriate many-body equations
                                                           in some approximation.
However, due to the computational difficulties and complications (e.g.,      the
  hard-core property in realistic nuclear force), in practical calculation, some
  simple effective interactions whose parameters are adjusted to reproduce gross
        nuclear properties rather than nucleon-nucleon scattering data, even the
         one-body potential, are usually used by combining a phenomenological or
                                             self-consistent mean-field approach.
Presently, there are four types of models which are widely used              for
 investigating fission barriers, including the macroscopic-microscopic      (MM)
             models~\cite{Moller2004,Sobiczewski2006,Dobrowolski2007,Moller2009,
       Dobrowolski2009,Kowal2010,Moller2015}, the nonrelativistic energy density
                   functionals based on zero-range Skyrme and finite-range Gogny
     interactions~\cite{Staszczak2009,Bender1998,Bonneau2004,Staszczak2007}, the
                          extended Thomas-Fermi plus Strutinsky integral (ETFSI)
                                    methods~\cite{Mamdouh2001,Dutta2000}, and the
          covariant density functional theory (CDFT)~\cite{Abusara2010,LiZP2010,
                                                           Ring2011,Abusara2012}.
The difference in the description of inner fission barrier height in       these
       models is considerable, which will translate into huge uncertainties in
                          the spontaneous fission half-lives~\cite{Agbemava2017}.
The MM approaches with the ``oldest'' ages usually have very     high
                          descriptive power as well as simplicity of calculation.
In this paper, the multidimensional potential energy surface (PES)    calculations
                                          are based on the framework of MM models.
%1234567890123456789012345678901234567890123456789012345678901234567890123456789

Experimental data show some unexpected spectroscopic characteristics     in
      some nuclei, such as wobbling, signature inversion (or splitting) and chiral
   doublets~\cite{Odegard2001,Bengtsson1984,Starosta2001}, indicating the possible
                                                   appearance of triaxiality.
Prior to this work, we have performed several studies of the effects          of
 triaxial deformation degree of freedom on ground and/or yrast state in specific
  isotopes such as Ba~\cite{YangJ2015}, Nd~\cite{CHAI2015} and W~\cite{YANG2015}
                                   with a similar PES calculation method.
Additionally, we have systematically investigated the octupole effects on  outer
                                      fission barriers for even-even nuclei with
                                  $102\leq Z\leq 112$~\cite{Wang2012CSB}.
In this work we perform a systematic investigation of the       inner
      fission barriers in all synthesized tranuranium nuclei within the triaxial
       multidimensional PES approach, focusing on the formation mechanism of the
       barriers and triaxiality effects, which is expected to provide a valuable
  reference for experiments, and to some extent even for more microscopic and relativistic calculations.
The barrier shape, including its height, width and evolution, is analyzed.
To our knowledge, such systematic investigations of the effects of       triaxial
  degrees of freedom on the height of inner fission barriers are somewhat scarce
   for all even-even transuranium nuclides ranging from $^{226}$Pu to $^{294}$Og
                                             which have already been synthesized
         experimentally~\cite{NNDC,Oganessian2002,Oganessian2007,Oganessian2017}.
Moreover, a systematic study may be the best way to understand the underlying principles behind the impact of
     triaxiality on the inner fission barriers, since it can eliminate the
   arbitrariness of conclusions with respect to the choice of a specific nucleus.
Part of the aim of this work is to test the model validity and predictive power,
         especially extrapolating towards the the superheavy region, and to find
                          discrepancies for further developing the present model.
Indeed, for instance, it has been pointed out that different polarization effects and
 functional forms of the densities may appear in the superheavy region, which can be
            naturally incorporated within the self-consistent nuclear mean-field
                                                                    calculations.
However, in the MM models, prior knowledge                       about
           the expected densities and single-particle potentials is needed~\cite{Rutz1997}.
In general, the model parameters may need to be refitted and even the      model
                               Hamiltonian will need to be remodelled (e.g., see
                                Refs.~\cite{Belgoumene1991,Dudek2010,Dudek2013}).

The rest of this paper is organized as                                   follows.
The theoretical framework and the details of the numerical calculations      are
                                                             described in Section~2.
The calculated fission barriers, the effect of triaxiality, and the   comparison
     with data and other theoretical results are presented             in Section~3.
Finally, Section~4 summarizes the main conclusions of the present                  work.

%1234567890123456789012345678901234567890123456789012345678901234567890123456789
\section{Theoretical framework}\label{sec.II}

The PES calculation~\cite{XuFR1998} employing in the present work is based on    the
                           MM models~\cite{Nix1972,Nazarewicz1989}, which are
                                   approximations of the self-consistent Hartree-Fock
                                          method~\cite{Strutinsky1967,Brack1972}.
Such approaches have been widely used to reproduce the right    bulk properties,
         e.g., the ground-state deformations and energies, of a many-body system
                      in medium and heavy mass nuclei~\cite{Regan2002,LiuHL2011}.
In this section, the common procedures are outlined, with some
                                                    helpful references.

The basic idea in the MM models is that the total potential energy of          a
                              deformed nucleus can be decomposed into two parts,
\begin{eqnarray}
     E_{\rm total}(Z,N,\hat{\beta})
     =
     E_{\rm mac} (Z,N,\hat{\beta})
     +
     E_{\rm mic} (Z,N,\hat{\beta}),
                                                                  \label{eqn.01}
\end{eqnarray}
where $E_{\rm mac}$ is the macroscopic energy given by a smooth      function of
              nucleon numbers ($Z$, $N$) and deformations ($\hat{\beta}$), while
   $E_{\rm mic}$ represents the microscopic quantum correction calculated from a
           phenomenological (non-self-consistent) single-particle potential well.
Generally, the unified procedure of such an approach is then carried       out   in
                                        the following five steps~\cite{Nix1972}:

       (a) Specify the nuclear shape, i.e., shape parameterization.               \\
\indent(b) Calculate the macroscopic energy, e.g., liquid-drop (LD) energy.       \\
\indent(c) Generate the single-particle potential felt by protons or neutrons,
           e.g., the Nilsson, Woods-Saxon (WS), and folded Yukawa (FY)
           potentials.                                                        \\
\indent(d) Solve the stationary Schr\"{o}dinger equation to obtain the
                                    single-particle levels and wave functions.\\
\indent(e) Calculate microscopic (shell and pairing) corrections.             \\
The total potential energy can obviously be obtained by the sum of   macroscopic
               and microscopic energies given in steps (b) and (e), respectively.
Next, each part will be introduced according to these        five          steps.

First of all, the nuclear shape can be conveniently described  by             the
      parametrization of the nuclear surface or the nucleon density distribution.
In the present work, the nuclear surface $\Sigma$ is depicted with           the
       multipole expansion of spherical harmonics $Y_{\lambda\mu}(\theta,\phi)$,
                                                                        that is,
%%%%%%%%%%%%%%%%%%%%%%%%%%%%%%%%%%%%%%%%%%%%%%%%%%%%%%%%%%%%%%%%%%%%%%%%%%%%%%%%
%123456789 123456789 123456789 123456789 123456789 123456789 123456789 123456789
\begin{equation}
    \Sigma:
    R(\theta,\phi)
    =
    R_0
    \Big[
    1
    +
    \sum_{\lambda}
    \sum_{\mu=-\lambda}^{+\lambda}
    \alpha_{\lambda\mu}
    Y^*_{\lambda\mu}(\theta,\phi)
    \Big],
                                                                  \label{eqn.02}
\end{equation}
%%%%%%%%%%%%%%%%%%%%%%%%%%%%%%%%%%%%%%%%%%%%%%%%%%%%%%%%%%%%%%%%%%%%%%%%%%%%%%%%
%123456789 123456789 123456789 123456789 123456789 123456789 123456789 123456789
where $R_0$ is the radius of the isovolume spherical shape, which is   determined
               by requiring conservation of the nuclear volume equal to
                $4\pi R_0^3/3$, and $\hat{\beta}$ stands for all the deformation
                                                         parameters applied here.
Note that such parametrization is also convenient to describe        the nuclear
                                                            geometrical symmetry.
Here, the dominating low-order quadrupole deformations            $(\alpha_{20},
                   \alpha_{2\pm2})$ and hexadecapole deformations $(\alpha_{40},
         \alpha_{4\pm2},\alpha_{4\pm4})$ have been included.
Meanwhile, the nuclear surface radius $R(\theta, \phi)$ represents  the distance
      of a point between the nuclear surface and the origin of the corresponding
                                                               coordinate system.
Since only the even $\lambda$ and $\mu$ components are included,     the nuclear
                      shape will obviously survive three symmetry planes by such
                                                                 parametrization.
Furthermore, once the hexadecapole deformation is taken into account with      the
       functions of the scalars in the quadrupole tensor $\alpha_{2\mu}$, it can
                    lead to a three-dimensional calculation with the independent
     coefficients $\beta_2$, $\gamma$ and $\beta_4$~\cite{Nazarewicz1985}, i.e.,
%%%%%%%%%%%%%%%%%%%%%%%%%%%%%%%%%%%%%%%%%%%%%%%%%%%%%%%%%%%%%%%%%%%%%%%%%%%%%%%%
%123456789 123456789 123456789 123456789 123456789 123456789 123456789 123456789
\begin{equation}
  \left\{
   \begin{array}{lcl}
     \alpha_{20}
    =
     \beta_2 \rm cos\gamma
       \\%[1mm]
     \alpha_{22}
    =
     \alpha_{2-2}
    =
     \frac{\sqrt{2}}{2}\beta_2 \rm sin\gamma
       \\%[1mm]
     \alpha_{40}
     =
     \frac{1}{6}\beta_4(5 \rm cos^2\gamma+1)
       \\%[1mm]
     \alpha_{42}
     =
     \alpha_{4-2}
     =
     -\frac{1}{12}\sqrt{30}\beta_4 \rm sin2\gamma
       \\%[1mm]
     \alpha_{44}
     =
     \alpha_{4-4}
     =
     \frac{1}{12}\sqrt{70}\beta_4 \rm sin^2\gamma.
       \\%[1mm]
   \end{array}
  \right.
                                                                  \label{eqn.03}
\end{equation}
%%%%%%%%%%%%%%%%%%%%%%%%%%%%%%%%%%%%%%%%%%%%%%%%%%%%%%%%%%%%%%%%%%%%%%%%%%%%%%%%
%123456789 123456789 123456789 123456789 123456789 123456789 123456789 123456789
Of course, the                 ($\beta_2,\gamma,\beta_4$)
                                                         parametrization has all
          the symmetry properties (e.g., axial symmetry and reflection symmetry)
           of Bohr's ($\beta_2,\gamma$) parametrization~\cite{Bohr1952,Bohr1998}.

Secondly, up to now, there are several phenomenological LD models which have been used  in the
              literature, e.g., standard LD model~\cite{Myers1966}, finite-range
      liquid-drop model (FRLDM)~\cite{Moller1988LDM}, finite-range droplet model
                        (FRDM)~\cite{Moller1988DM}, Lublin-Strasbourg drop (LSD)
                                                  model~\cite{Pomorski2003}, etc.
These macroscopic models with slightly different properties can be utilized   to
              give the smoothly varying part of the nuclear energy, in which the
            dominating terms involve the volume energy, the surface energy and
                                                              the Coulomb energy.
Due to the incompressibility of a nucleus (``volume conservation''   condition),
   the volume energy, which is proportional to mass number $A$, does not depend
       on the nuclear shape, whereas the surface energy, which tends to hold the
       nucleus together, and the Coulomb energy, which tends to pull the nucleus
                                                      apart, are shape-dependent.
Here, the macroscopic energy calculated in the present work is obtained   by the
                   standard LD model with the parameter set used by Myers and
                                                      Swiatecki~\cite{Myers1966}.
Because we are focusing on the PES and the difference               between  the
               critical points (e.g., between the minimum and saddle point), the
      potential energy relative to the energy of a spherical LD has been adopted.
This portion of the standard LD energy can be can                     be written
                                    as~\cite{Cwiok1987, Myers1966,Bolsterli1972}
\begin{equation}
E_{\rm LD}(Z,N,\hat{\beta})
=
\{[B_s(\hat{\beta})-1]+2\chi[B_c(\hat{\beta})-1]\}E_s^{(0)},
                                                                  \label{eqn.04}
\end{equation}
where the relative surface energy $B_s$ and Coulomb energy $B_c$ are    functions
                  only of nuclear shape, depending on the collective coordinates
     $\{\alpha_{\lambda\mu}\}$. The spherical surface energy $E_s^{(0)}$ and the
                           fissility parameter $\chi$ are $Z$- and $N$-dependent.
The detailed expression of  $E_s^{(0)}$ and $\chi$ can be               found in
                                                       Ref.~\cite{Bolsterli1972}.

Thirdly and fourthly, in the one-body mean-field approximation,     for example,
            the Nilsson (modified harmonic-oscillator), WS and FY potentials are
                                      usually adopted during the process of real
                            calculations~\cite{Nix1972,Bengtsson1989,Moller1995}.
At present, we use a more realistic diffuse-surface deformed WS-type     nuclear
                                                                       potential.
That is, the single-particle levels and wave functions are determined         by
                solving  the stationary Schr\"{o}dinger equation numerically with
                                                  $H_{\rm WS}$~\cite{Dudek1980},
%%%%%%%%%%%%%%%%%%%%%%%%%%%%%%%%%%%%%%%%%%%%%%%%%%%%%%%%%%%%%%%%%%%%%%%%%%%%%%%%
%123456789 123456789 123456789 123456789 123456789 123456789 123456789 123456789
\begin{eqnarray}
    H_{\rm WS}
    &=&
    -\frac{\hbar^2}{2m}\nabla^2
    +
    V_{\rm cent}(\vec{r};\hat{\beta})
    +
    V_{\rm so}(\vec{r},\vec{p},\vec{s};\hat{\beta})
    \nonumber\\
    &&+
    \frac{1}{2}(1+\tau_3)V_{\rm Coul}(\vec{r},\hat{\beta}),
    \nonumber\\
                                                                  \label{eqn.05}
\end{eqnarray}
%%%%%%%%%%%%%%%%%%%%%%%%%%%%%%%%%%%%%%%%%%%%%%%%%%%%%%%%%%%%%%%%%%%%%%%%%%%%%%%%
%123456789 123456789 123456789 123456789 123456789 123456789 123456789 123456789
where the depth of the central part of the WS potential, which       will mainly
         govern the number of levels in the potential well, is~\cite{Cwiok1987},
%%%%%%%%%%%%%%%%%%%%%%%%%%%%%%%%%%%%%%%%%%%%%%%%%%%%%%%%%%%%%%%%%%%%%%%%%%%%%%%%
%123456789 123456789 123456789 123456789 123456789 123456789 123456789 123456789
\begin{equation}
    V
    =
    V_0[1\pm\kappa(N-Z)/(N+Z)],
                                                                  \label{eqn.06}
\end{equation}
%%%%%%%%%%%%%%%%%%%%%%%%%%%%%%%%%%%%%%%%%%%%%%%%%%%%%%%%%%%%%%%%%%%%%%%%%%%%%%%%
%123456789 123456789 123456789 123456789 123456789 123456789 123456789 123456789
with the plus and minus signs  for protons and neutrons respectively,    and
                  the values of the constants $V_0$ and $\kappa$ given later.
Then the spin-orbit potential  $V_{\rm so}(\vec{r},\vec{p},\vec{s};\hat{\beta})$,
                   which mainly controls the relative positions of levels, is
                                                             assumed of the form
%%%%%%%%%%%%%%%%%%%%%%%%%%%%%%%%%%%%%%%%%%%%%%%%%%%%%%%%%%%%%%%%%%%%%%%%%%%%%%%%
%123456789 123456789 123456789 123456789 123456789 123456789 123456789 123456789
\begin{eqnarray}
    V_{\rm so}(\vec{r},\vec{p},\vec{s};\hat{\beta})
    &=&
    -\lambda
    \Big[
    \frac{\hbar}{2mc}
    \Big]^2
    \nonumber\\&&
    \bigg \{
    \nabla\frac{V_0[1\pm\kappa(N-Z)/(N+Z)]}{1
    +
    \textrm{exp}[\textrm{dist}_{\Sigma_{\rm{so}}}
    (\vec{r},\hat{\beta})/a_{\rm{so}}]}
    \bigg \}
    \times\vec{p}\cdot\vec{s},
    \nonumber\\
                                                                  \label{eqn.07}
\end{eqnarray}
%%%%%%%%%%%%%%%%%%%%%%%%%%%%%%%%%%%%%%%%%%%%%%%%%%%%%%%%%%%%%%%%%%%%%%%%%%%%%%%%
%123456789 123456789 123456789 123456789 123456789 123456789 123456789 123456789
where $\lambda$ is the strength parameter of the spin-orbit potential and    the
     new surface $\Sigma_{\rm so}$ denotes the surface of the  spin-orbit potential.
In addition, the Coulomb potential $V_{\rm Coul}(\vec{r},\hat{\beta})$       for
               protons is considered as a uniformly charged drop in a classical
                                                         electrostatic potential.

The universal WS parameter    set~\cite{Cwiok1987}
                                            has been applied in the present work.
This parameter set is $Z$ and $N$ independent and can         give a reliable
                    description of the single-particle states, especially in the
                                medium and heavy mass regions~\cite{Leander1984}.
As shown in Refs.~\cite{Cwiok1987,Nazarewicz1989}, these parameters          are:

       (1) Central potential depth parameters:                                \\
\indent\indent $V_0=49.6$ MeV, $\kappa=0.86$.                                 \\
\indent        (2) Radius parameters of the central part:                     \\
\indent\indent $r_0(p)=1.275$ fm, $r_0(n)=1.347$ fm.                          \\
\indent        (3) Radius parameters of the spin-orbit part:                  \\
\indent\indent $r_{0-\rm so}(p)=1.320$ fm, $r_{0-\rm so}(n)=1.310$ fm.        \\
\indent(4) Strength of the spin-orbit potential:                              \\
\indent\indent $\lambda(p)=36.0$, $\lambda(n)=35.0$.                          \\
\indent(5) Diffuseness parameters:                                            \\
\indent\indent $a_0(p)=a_0(n)=a_{0-\rm so}(p)=a_{0-\rm so}(n)=0.70$ fm.       \\
These are usually fitted by adjusting their                     values to
  optimally reproduce the energies of available single-particle levels in either
   spherical or deformed nuclei~\cite{Dudek1978,Dudek1979,Dudek1981,Nojarov1984}.
It is worth noting that these parameters may be not      constant
    throughout the global nuclear chart and need to be adjusted somehow
                                        when extrapolating to the unknown region.
Based on the above parameters, the WS single-particle potential          felt by
              protons and neutrons is generated clearly at the deformation space
                                                      ($\beta_2,\gamma,\beta_4$).
Then the WS Hamiltonian matrix is calculated using the basis of the   axially
               deformed harmonic oscillator in the cylindrical coordinate system.
Finally, the single-particle levels and wave functions can be obtained         by
                                            diagonalizing the Hamiltonian matrix.
Note that the harmonic oscillator eigenfunctions with the principal      quantum
    number $N\leq 12$ and $N\leq 14$ are  adopted during the calculation as a basis for protons and
       neutrons respectively, since such a basis cutoff can give stable
                             results in the case of its possible enlargement.

Fifthly, the microscopic energy arises as a result of the          inhomogeneous
           distribution of the single-particle energies and the residual pairing
             interaction in the nucleus, mainly consisting of a shell correction
  $\delta E_{\rm shell}$ and a pairing correction $\delta E_{\rm pair}$, namely,
\begin{eqnarray}
     E_{\rm mic} (Z,N,\hat{\beta})
     =
     \delta
     E_{\rm shell} (Z,N,\hat{\beta})
     +
     \delta
     E_{\rm pair} (Z,N,\hat{\beta}).
                                                                  \label{eqn.08}
\end{eqnarray}
Based on the single-particle levels obtained above, the shell and        pairing
         corrections at each deformation point ($\beta_2,\gamma,\beta_4$) can be
           evaluated by means of the Strutinsky method~\cite{Strutinsky1967} and
                                    Lipkin-Nogami (LN) method~\cite{Pradhan1973}.
The Strutinsky smoothly varying part is carried out with a six-order    Laguerre
              polynomial with the smoothing range $\gamma=1.20$ $\hbar\omega_0$
                                                ($\hbar\omega_0=41/A^{1/3}$ MeV).
The LN method, in which particle number projection is approximately       conserved,
                 avoids the spurious pairing phase transition encountered in the
                         traditional Bardeen-Cooper-Schrieffer (BCS) calculation.
The monopole pairing has been considered and its strength, $G$, is obtained from
                                        the average gap method~\cite{Moller1992}.
In the pairing windows, the respective states, e.g., half of the particle number
 $Z$ and $N$ (or 40, if they are greater than 40) just below and above the Fermi
                  energy, are included empirically for both protons and neutrons.
Then the energy in the LN approach is given by~\cite{Pradhan1973,Moller1995},
\begin{eqnarray}
     E_{\rm LN}
     =
     \sum_k
     2v_k^2 e_k
     -
     \frac{\Delta^2}{G}
     -
     G
     \sum_k
     v_k^4
     +
     G
     \frac{N}{2}
     -
     4\lambda_2
     \sum_k
     u_k^2
     v_k^2,
%     \nonumber\\
                                                                  \label{eqn.09}
\end{eqnarray}
where $v_k^2$, $e_k$ and $\Delta$ denote                         the
              occupation probabilities, single-particle energies and pairing gap respectively.
The extra Lagrange multiplier $\lambda_2$ represents                         the
                                            particle-number-fluctuation constant.
Further, the shell and pairing corrections could be                calculated by
                     $\delta E_{\rm shell}=\sum e_i - \tilde{E}_{\rm Strut}$ and
                     $\delta E_{\rm pair} =E_{\rm LN} - \sum e_i $, respectively.
Here, $\sum e_i$ is the sum of single-particle energies                      and
            $\tilde{E}_{\rm Strut}$ is the smoothing energy by the Strutinsky method.

Last but not least, taking both Bohr                           shape deformation
    parameters~\cite{Bohr1952} and the Lund convention~\cite{Andersson1976} into
                                   account, the Cartesian quadrupole coordinates
     $X=\beta_2\rm cos(\gamma+30^\circ)$ and $Y=\beta_2\rm sin(\gamma+30^\circ)$
                                             are used in the present work,
                       where $\beta_2$ specifies the magnitude of the quadrupole
                              deformation and $\gamma$ describes nonaxial shapes.
In these calculations, the $\beta_2$ value ``built-in'' is always positive   and
      the $\gamma$ value covers the range $-120^\circ \leq \gamma \leq 60^\circ$.
Obviously, the three sectors $[-120^\circ, -60^\circ]$,   $[-60^\circ, 0^\circ]$
    and $[0^\circ, 60^\circ]$ describe the identical triaxial shapes at the
                                                                    ground state.
At each ($X, Y$) deformation grid, the total energy of a nucleus   is calculated
   according to the procedure mentioned above and the PES can finally be derived
      from interpolating, using a cubic spline function, between the lattice
                                                    points in the ($X, Y$) plane.
Therefore, the nuclear properties such as the ground-state           equilibrium
        deformations, saddle points, fission paths and so on can be obtained and
                                   analyzed based on the present PES calculation.
%1234567890123456789012345678901234567890123456789012345678901234567890123456789
\section{Results and discussion}\label{sec.III}
\end{multicols}
\vspace{-6mm}
\begin{center}
\includegraphics[width=7.5cm]{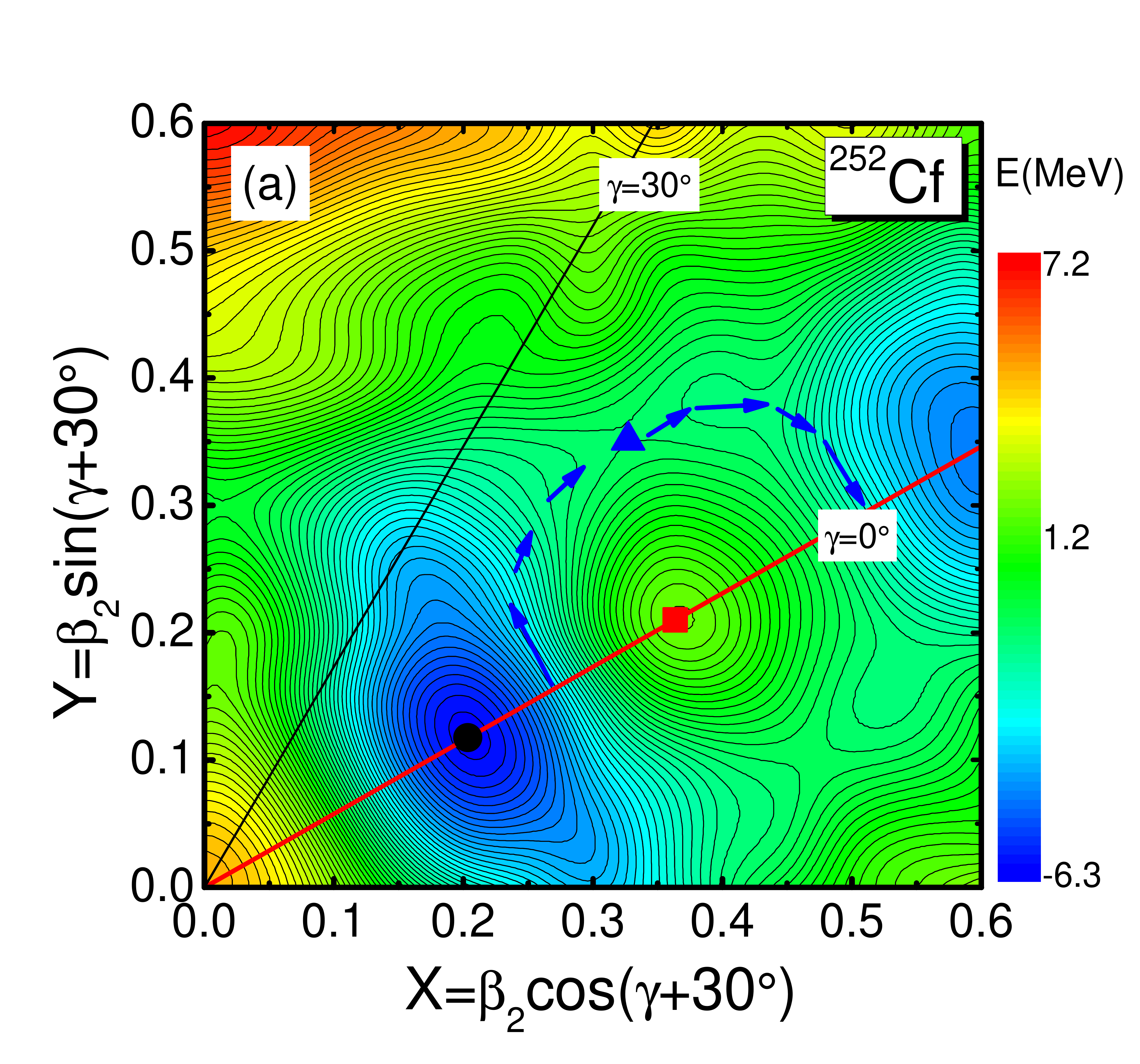}
\includegraphics[width=7.5cm]{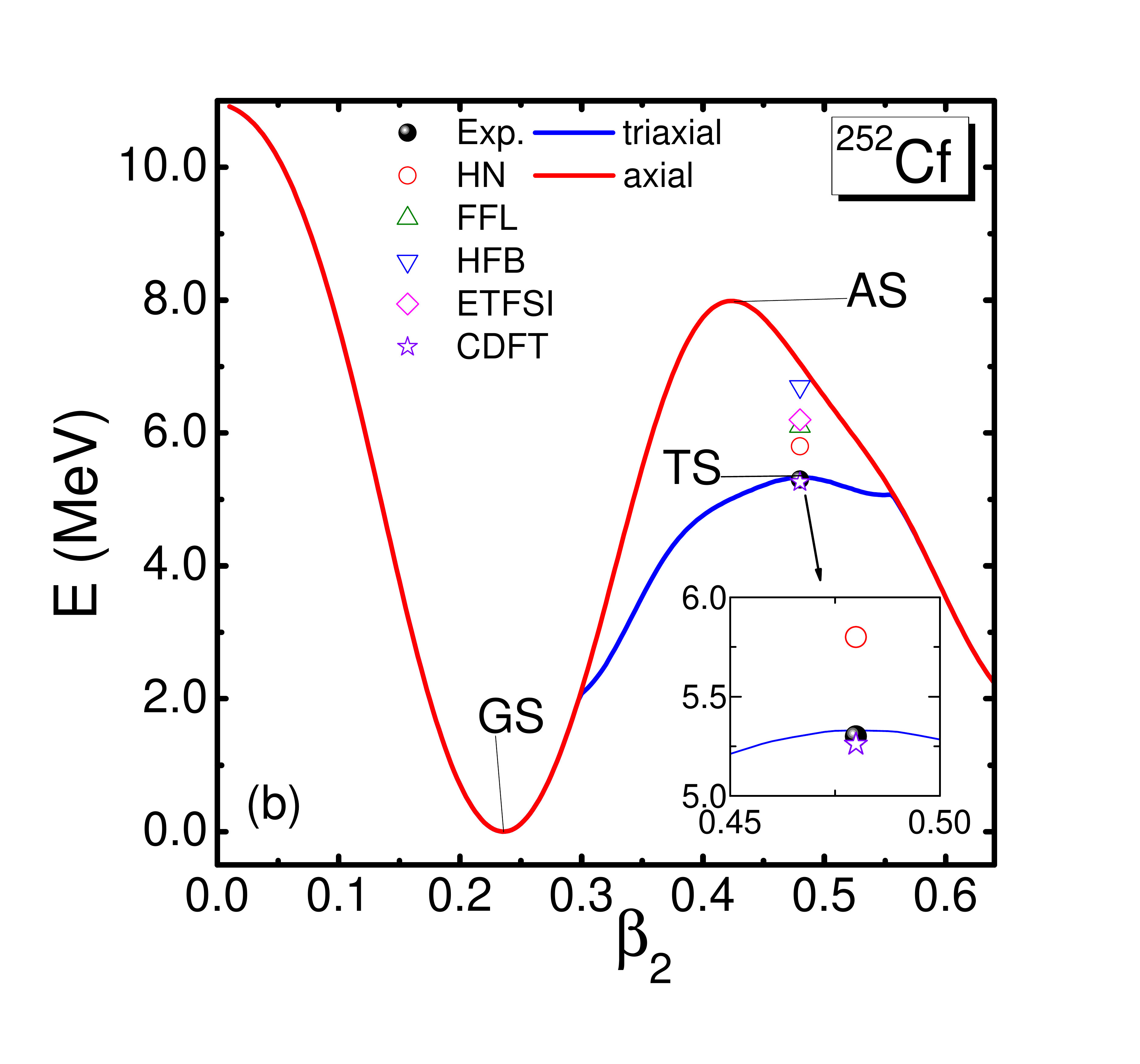}
\vspace{-3mm}
\figcaption{
(a) Calculated two-dimensional PES in           ($\beta_2$, $\gamma$, $\beta_4$)
                           deformation space for the selected nucleus $^{252}$Cf.
At each ($\beta_2$, $\gamma$) grid point, the PES has been minimized        with
                                  respect to the deformation parameter $\beta_4$.
The PES minimum, namely, the ground state (GS), is indicated by the black     circle.
The red square and blue triangle represent the axial saddle (AS) and    triaxial
  saddle (TS) along the axial (red line) and triaxial (blue line) fission paths,
                                                                    respectively.
The energy contours are at 200 keV                                     intervals.
(b) Calculated potential energy curves against $\beta_2$ for          $^{252}$Cf.
At each given $\beta_2$ point, the energy has been minimized with respect     to
                                             $\gamma$ and $\beta_4$ deformations.
Similar to (a), the red solid line displays the potential energy curve for   the
       axially symmetric solution with $\gamma=0^\circ$, whereas the blue solid line
 shows the corresponding curve along the triaxial part of the fission trajectory
                          which has the lowest energy as a function of $\beta_2$.
For convenience of description, the energy curve is normalized with respect   to
                                                         the ground-state energy.
The experimental and other several theoretical fission barriers are shown    for
                                                                      comparison.
Further details are given in the                                            text.
                                                                   \label{FIG.1}
}
\end{center}
\begin{multicols}{2}
%%%%%%%%%%%%%%%%%%%%%%%%%%%%%%%%%%%%%%%%%%%%%%%%%%%%%%%%%%%%%%%%%%%%%%%%%%%%%%%%
%description for Fig.1

In the present work, we just care about the inner fission barriers rather than the
                                                                      outer ones.
Such a restriction has its own                                     merits.
The inner barriers are easier to measure than the outer barriers, and
they are more important for the $r$ process since they
                                                            determine thresholds.
In addition, spontaneous fission lifetimes tend to be dominated by the     inner
   barrier, even if an outer barrier can occasionally have a crucial influence if it
                                                                  is wide enough.
Previous studies~\cite{Abusara2010,Sobiczewski2006,Staszczak2007,      Rutz1995,
                          Cwiok1996,Samyn2005} have shown that the odd-multipole
          deformations (e.g., $\beta_3$) do not play a role in the inner fission
                                  barrier of the actinide and superheavy regions.
This allows us to restrict our calculations to reflection symmetric       shapes.
It has also been shown that the inclusion of triaxiality can improve     the
    accuracy of the description of the inner fission barriers in the actinide region
                                                  in almost all state-of-the-art
     models~\cite{Abusara2012,LuBN2014,Moller2009,Dobrowolski2007,Delaroche2006}.
Therefore, taking the important low-order deformation degrees of freedom    into
       account, we performed the PES calculation in multidimensional ($\beta_2$,
   $\gamma$, $\beta_4$) space, paying attention to the impact of triaxiality  on the
                      inner fission barrier, especially in the superheavy nuclei.
To examine the model validity in the current research, as an example, the $^{252}$Cf
      nucleus is calculated, as it is the closest to the superheavy region among the
                        nuclei that have experimentally measured inner fission barriers~\cite{IAEA1993,RIPL2009}.
Figure~\ref{FIG.1} shows the calculated 2D PES (a) and the triaxial and axial 1D
                                      potential energy curves (b) for $^{252}$Cf.
The axial and triaxial fission paths, saddles and      equilibrium
                         deformations can  be clearly seen in Fig.~\ref{FIG.1}(a).
For instance, one can see that this nucleus has a well-deformed prolate    shape
       and a triaxial-deformed saddle, $\gamma\sim 17^\circ$, which will strongly
                                                         modify the fission path.
The ``real'' fission path will go through the energy valley (denoted by the blue
   line) with a non-zero $\gamma$ deformation in the ($\beta_2$, $\gamma$) plane
                            by avoiding the maximum of the axially symmetric PES.
The properties of the inner fission barrier, including its height and width,   are
                                    shown in Fig.~\ref{FIG.1}(b).
The triaxiality lowers the barrier height of this nucleus by up to 2.66 MeV here.
For comparison, the experimental data and other several theoretical results   of
               the inner fission barrier for $^{252}$Cf are also shown.
Note that these several theoretical results are, respectively, obtained     from
  the so-called heavy nuclei (HN) model~\cite{Kowal2010}, the FY single-particle
                                      potential and the FRLDM~\cite{Moller2009},
    the Skyrme-Hartree-Fock-Bogoliubov (SHFB) method~\cite{Samyn2005}, the ETFSI
                  methods~\cite{Mamdouh2001} and the CDFT theory~\cite{LuBN2014}.
It can be seen that all the calculated values are higher than the data    except
                                  for the CDFT value, which has a 40~keV underestimation.
Interestingly, our result shows good agreement with experimental  data,
                   at least in this nucleus, with only  a $\sim$30~keV difference.
%%%%%%%%%%%%%%%%%%%%%%%%%%%%%%%%%%%%%%%%%%%%%%%%%%%%%%%%%%%%%%%%%%%%%%%%%%%%%%%%
%description for Fig.2

Nowadays, it is well understood that the minima (ground or      shape-coexisting
     states) and maxima (or saddle points) can be attributed to shell effects,
  whose microscopic mechanism originates from the nonuniform distribution of the
                             single-particle levels in the vicinity of the Fermi
                                                   surface~\cite{Strutinsky1967}.
Minima correspond to regions of low level density, e.g., a region with  a  large
        energy gap, whereas saddle points usually occur in the vicinity of level
                                     crossings, regions of high level density.
The fission barrier (namely, the energy difference between the      ground-state
   minimum and the corresponding saddle point) is certainly related to the level
                                                    density near the Fermi level.
For additional clarity and emphasis, Fig.~\ref{FIG.2} shows calculated  proton
 and neutron single-particle levels near the Fermi surface at three      typical
       deformation points (ground-state minimum, axial and triaxial
                                    saddles) for the selected $^{252}$Cf nucleus.
The single-particle level density near the Fermi level at   the
         ground-state minimum is lower than those at the saddles, indicating a large
                                                negative shell correction energy.
Especially, it is very clear that the highest neutron level     density is
 near the Fermi level at the axial saddle point, as seen in Fig.~\ref{FIG.2}(b),
       since at both the ground-state minimum and triaxial saddle the levels are
   shifted up above the Fermi level and down below the Fermi level (leading to a
                                                 relatively lower level density).
The decrease of the triaxial saddle (or the lowering of the barrier due       to
           triaxiality) can certainly be understood by such a level density change.
%%%%%%%%%%%%%%%%%%%%%%%%%%%%%%%%%%%%%%%%%%%%%%%%%%%%%%%%%%%%%%%%%%%%%%%%%%%%%%%%
\begin{center}
\includegraphics[width=7.5cm]{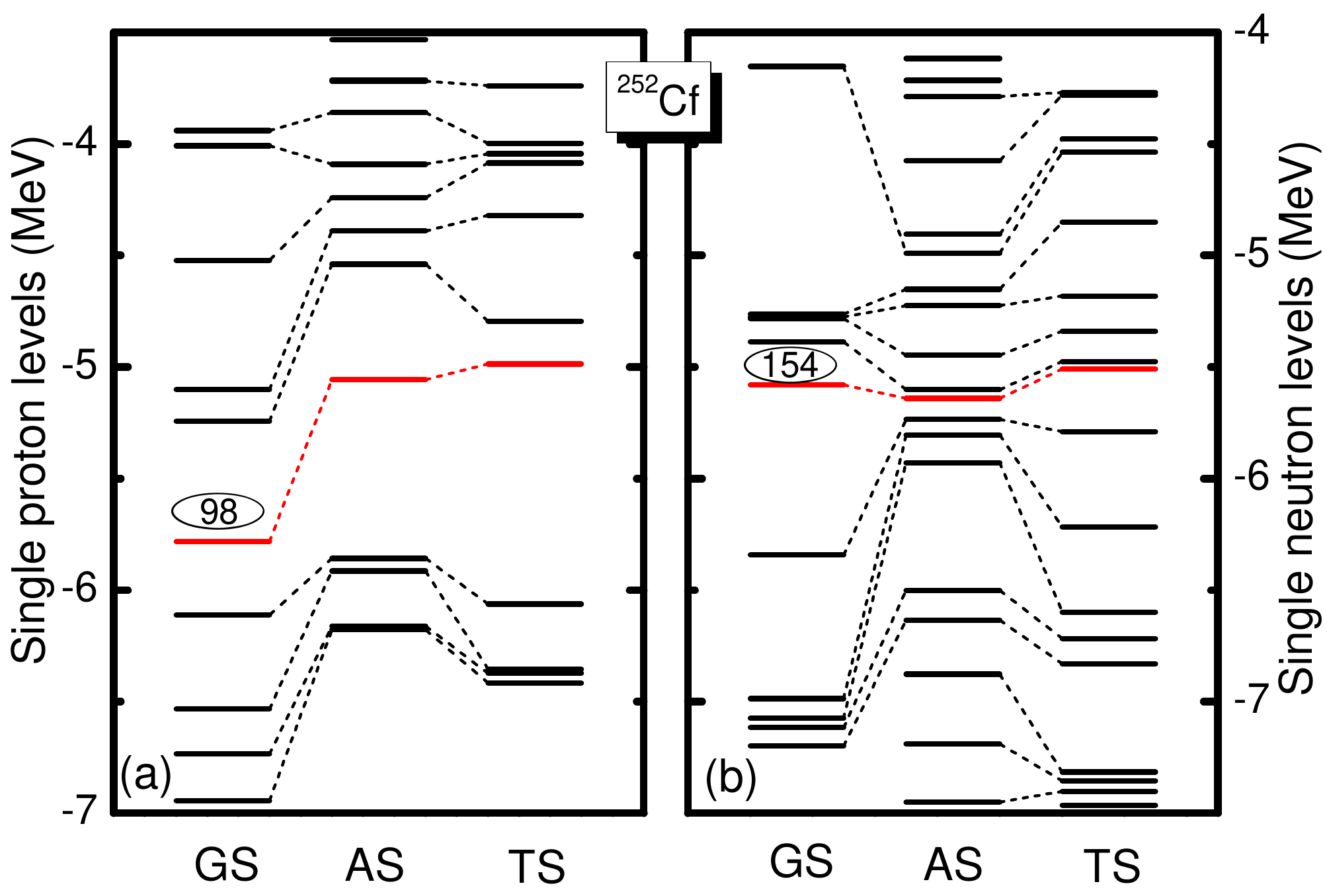}
\figcaption{
Calculated proton (a) and neutron (b) single-particle levels near the      Fermi
 surface for $^{252}$Cf at three typical deformation grid points, that is,   GS,
                                   AS and TS points, as seen in Fig.~\ref{FIG.1}.
The red lines indicate the Fermi energy                                   levels.
                                                                   \label{FIG.2}
}
\end{center}
%%%%%%%%%%%%%%%%%%%%%%%%%%%%%%%%%%%%%%%%%%%%%%%%%%%%%%%%%%%%%%%%%%%%%%%%%%%%%%%%
%description for Fig.3

To give a better understanding of the macroscopic and microscopic  contributions
         in the calculated PES, especially at critical minima and saddle points,
             Fig.~\ref{FIG.3} shows the calculated potential energy curves and
                                         corresponding histograms for $^{252}$Cf.
It can be seen from Fig.~\ref{FIG.3}(a) that the inclusion of triaxiality   does
   not change the first minimum but strongly affects the shape and height of the
                                                                   first barrier.
The macroscopic energy changes smoothly and the triaxiality will lead to       an
                                                      additional energy increase.
However, the microscopic energy fluctuates significantly and can be strongly affected  by
  the triaxiality, mostly determining the positions of the minima and
                                                                         saddles.
Note that at the zero deformation point, the reason for $E_{\rm macro}=0$ is  that
 the macroscopic energy is normalized to the spherical liquid drop, as mentioned
                                                         in the theoretical part.
More intuitively, Fig.~\ref{FIG.3}(b) shows us to what extent the  energy
           of the minimum and saddles are affected by the macroscopic energy and
                                                          microscopic correction.
%%%%%%%%%%%%%%%%%%%%%%%%%%%%%%%%%%%%%%%%%%%%%%%%%%%%%%%%%%%%%%%%%%%%%%%%%%%%%%%%
\begin{center}
\includegraphics[width=6cm]{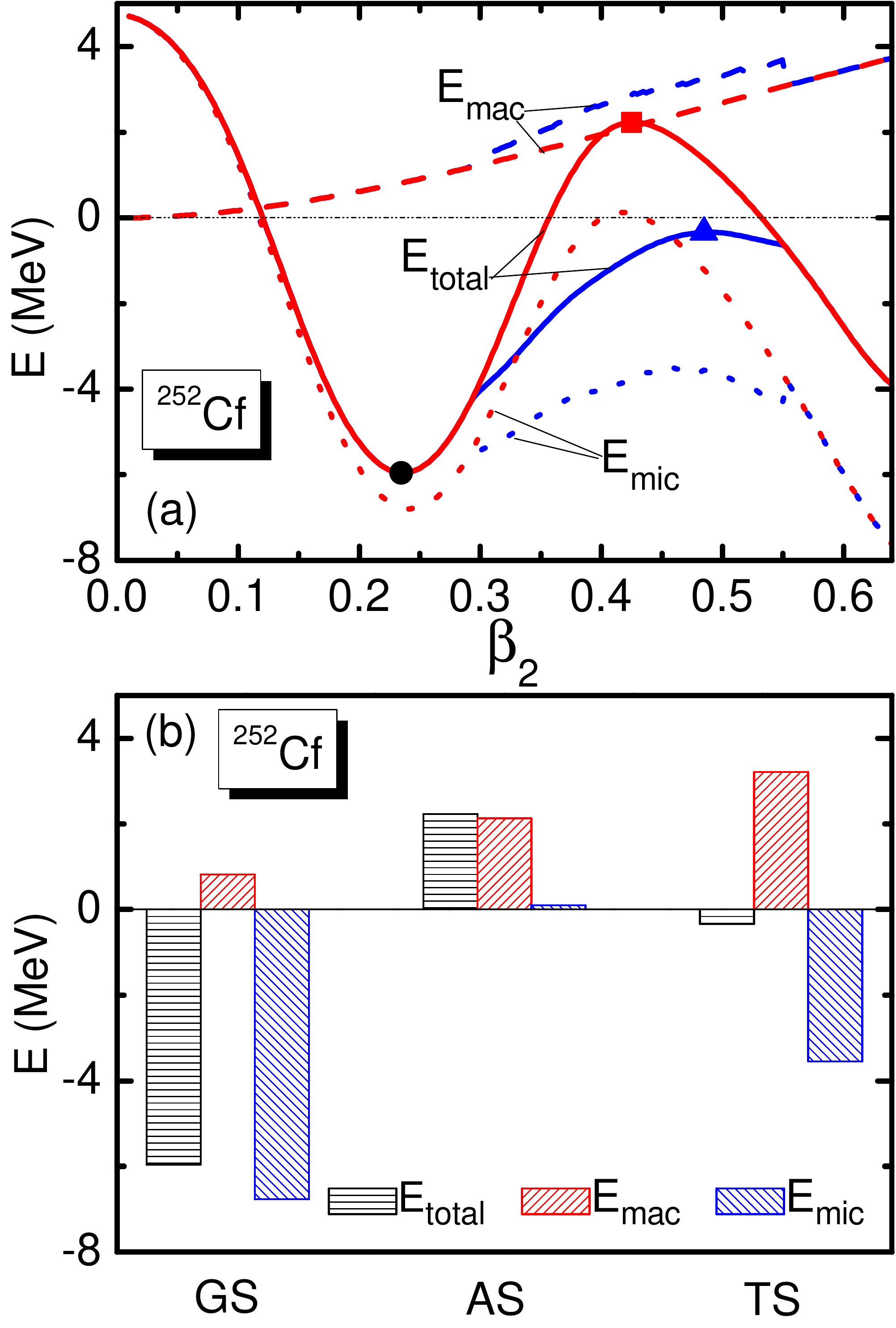}
\figcaption{
(a) Similar to Fig.~\ref{FIG.1}(b), calculated potential energy curves   against
     $\beta_2$ for $^{252}$Cf, but together with the macroscopic and microscopic
                                                                      components.
The circle, triangle and square symbols represent the ground state         (GS),
                        axial saddle (AS) and triaxial saddle (TS), respectively.
Note that, for simplicity, the $\beta_4$ deformation is not considered      here.
(b) The calculated total energy and its macroscopic and              microscopic
                           contributions at the GS, AS and TS points of part (a).
                                                                   \label{FIG.3}
}
\end{center}
%%%%%%%%%%%%%%%%%%%%%%%%%%%%%%%%%%%%%%%%%%%%%%%%%%%%%%%%%%%%%%%%%%%%%%%%%%%%%%%%
%description for Fig.4

Aside from the mean field, which determines the single-particle levels,       the
      pairing correlations can also affect the calculated barrier to some extent.
For example, it has been pointed out that there may be a rather large difference in         the
                              predicted barrier heights between different pairing
                          models~\cite{Karatzikos2010, Abusara2010, Abusara2012}.
In addition, with the dynamical coupling between shape and pairing    degrees of
 freedom, the fission barrier also will be reduced by several units of least-action
                                             of fission paths~\cite{Zhaojie2016}.
Here, to obtain a crude estimation of the contribution due to   the pairing
    correlation, as shown in Fig.~\ref{FIG.4}, the microscopic energy correction
                     has been further divided into shell and pairing corrections.
One can see that the $\beta_4$ deformation can slightly modify the     different
                                                               energy components.
As expected, it is found that the shell corrections are fully in agreement  with
              the cases of the density distribution, as seen in Fig.~\ref{FIG.2}.
For instance, the ground-state minimum has the lowest level       density,
         corresponding to the largest negative shell correction and the smallest
    negative pairing correction, whereas the axial saddle point has the highest
     level density, corresponding to a positive shell correction and the largest
                                                            negative pairing correction.
It seems that the triaxiality will change the single-particle levels by changing
  the phenomenological mean field, then change the shell and pairing corrections
                               and finally have an effect on the fission barrier.
%%%%%%%%%%%%%%%%%%%%%%%%%%%%%%%%%%%%%%%%%%%%%%%%%%%%%%%%%%%%%%%%%%%%%%%%%%%%%%%%

\begin{center}
\includegraphics[width=7.5cm]{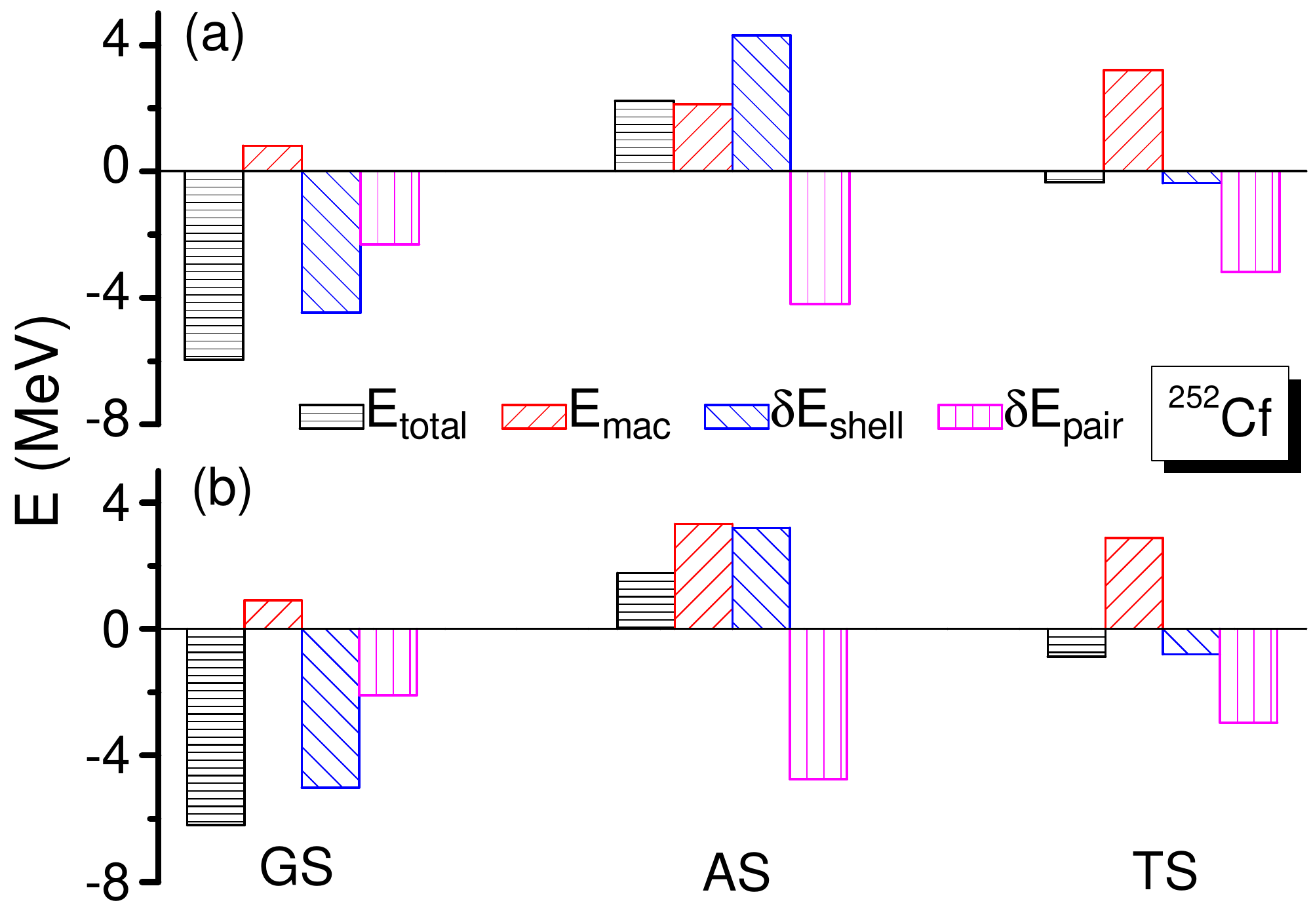}
\figcaption{
(a) Similar to Fig.~\ref{FIG.3}(b), but the microscopic energy is        further
                                      divided into shell and pairing corrections.
(b) The same as in (a), but the $\beta_4$ deformation is included in         the
                                                                     calculation.
                                                                   \label{FIG.4}
}
\end{center}
%%%%%%%%%%%%%%%%%%%%%%%%%%%%%%%%%%%%%%%%%%%%%%%%%%%%%%%%%%%%%%%%%%%%%%%%%%%%%%%%
%description for table.1
%123456789 123456789 123456789 123456789 123456789 123456789 123456789 123456789

As a basic rule of scientific research, theoretical quantities     generally
    need to be confronted with the corresponding experimental data and/or other
                                                               accepted theories.
Table~\ref{table} shows such two comparable quantities, the           calculated
      ground-state equilibrium deformation parameter $\beta_2$ and inner fission
    barriers $B_f$, for 13 even-even transuranium nuclei where the inner fission
                                    barriers have been determined experimentally.
The experimental $\beta_2$ values are deduced from the       intrinsic
         quadrupole moment related to the reduced electric quadrupole transition
   probability $B(E2)$~\cite{Pritychenko2016}. The other theoretical values
                               are, respectively, obtained from the so-called HN
                  model~\cite{Sobiczewski2001}, the FY single-particle potential
                                             and the FRDM~\cite{Moller2016}, the
                      Hartree-Fock-BCS (HFBCS)~\cite{Goriely2001}, and the ETFSI
                                                     methods~\cite{Aboussir1995}.
For the inner fission barriers, the experimental or empirical values are   taken
   from Refs.~\cite{IAEA1993,RIPL2009} and the other theoretical values come from
          the FY single-particle potential and the FRLDM~\cite{Moller2009},  the
                                            SHFB method~\cite{Samyn2005} and the
                                   CDFT~\cite{LuBN2014} in addition to the above
              mentioned HN~\cite{Kowal2010} and ETFSI~\cite{Mamdouh2001} methods.
It is clearly seen that all calculated $\beta_2$ values are lower           than
        the experimental results. The barriers calculated by different theories
are unevenly  distributed on both sides of the experimental values.
None of the theories can completely reproduce the  experimental data or even, as seen in the $B_f$ column,  always come  close to the experimental data.
It is hard to absolutely say which one is the best, in particular when extending  to
                                                               unknown nuclei.
However, it seems that these theories can to a large extent reproduce        the
                  deformed shapes and barrier amplitudes and support one another.
%%%%%%%%%%%%%%%%%%%%%%%%%%%%%%%%%%%%%%%%%%%%%%%%%%%%%%%%%%%%%%%%%%%%%%%%%%%%%%%%
%1234567890123456789012345678901234567890123456789012345678901234567890123456789
\end{multicols}
\begin{center}\scriptsize
\tabcaption{
The calculated results (PES) for ground-state            equilibrium deformation
           parameter $\beta_2$ and inner fission barriers $B_f$ for 13 available
                                                       even-even actinide nuclei.
The $\beta_2$ values of the HN~\cite{Sobiczewski2001},                   FY+FRDM
(FFD)~\cite{Moller2016}, HFBCS~\cite{Goriely2001}, and ETFSI~\cite{Aboussir1995}
 calculations and experiments (Exp.)~\cite{Pritychenko2016}, and the $B_f$ inner
  fission barriers of the HN~\cite{Kowal2010}, FY+FRLDM (FFL)~\cite{Moller2009},
           SHFB~\cite{Samyn2005}, ETFSI~\cite{Mamdouh2001}, CDFT~\cite{LuBN2014}
      calculations and experiments (Exp.)~\cite{IAEA1993,RIPL2009} are given for
                                                                      comparison.
                                                                   \label{table}
}
\begin{tabular}{p{1.2cm}ccccccc p{0.9cm}  ccccccc}
\hline \hline
\specialrule{0em}{1.5pt}{0pt} \small
{Nuclei}&\multicolumn{6}{c}{$\beta_2$}& & \multicolumn{7}{c}{$B_f$ (MeV)}     \\
\cline{2-7}  \cline{9-15}
&PES & HN  &FFD&HFBCS     &ETFSI
                       &Exp.$^a$
                       &&PES   &HN        &FFL    &SHFB &ETFSI &CDFT &Exp.    \\
   \hline
\specialrule{0em}{1.5pt}{0pt}
$^{236}$Pu  &0.215   &0.215   &0.215&0.26   &0.22     &---
                          &&\textbf{\emph{5.72}}$^b$ &5.4    &4.5
                          &4.7    &4.8   &---    &5.70                        \\
$^{238}$Pu  &0.220   &0.223   &0.226&0.24   &0.24     &0.282
                          &&6.32    &6.1    &5.3
                          &\textbf{\emph{5.4}}
                          &\textbf{\emph{5.4}}
                          &5.96   &5.60                                       \\
$^{240}$Pu  &0.225   &0.231   &0.237&0.25   &0.24     &0.290
                          &&6.48    &6.4    &\textbf{\emph{6.0}}
                          &5.9    &5.8   &5.92   &6.05                        \\
$^{242}$Pu  &0.227   &0.233   &0.237&0.24   &0.26     &0.298
                          &&6.38    &6.3    &6.4
                          &6.3    &6.2   &\textbf{\emph{5.77}}     &5.85      \\
$^{244}$Pu  &0.229   &0.235   &0.237&0.23   &0.26     &0.292
                          &&6.16    &\textbf{\emph{6.0}}
                          &6.6&6.5    &6.4   &\textbf{\emph{5.40}} &5.70      \\
$^{246}$Pu  &0.233   &0.239   &0.250&0.25   &0.26     &---
                          &&5.76    &\textbf{\emph{5.7}}
                          &6.3&6.5    &6.2   &4.76   &5.40                    \\
$^{242}$Cm  &0.229   &0.235   &0.237&0.25   &0.26     &---
                          &&6.59    &\textbf{\emph{6.7}}
                          &6.6&6.0    &6.1   &6.49   &6.65                    \\
$^{244}$Cm  &0.231   &0.237   &0.249&0.25   &0.26     &0.296
                          &&6.49    &6.6    &6.9&6.4    &6.4
                          &\textbf{\emph{6.34}}                    &6.18      \\
$^{246}$Cm  &0.233   &0.240   &0.249&0.27   &0.26     &0.298
                          &&6.29    &6.2    &7.0&6.7    &6.5
                          &\textbf{\emph{5.84}}                    &6.00      \\
$^{248}$Cm  &0.236   &0.242   &0.250&0.28   &0.26     &0.286
                          &&\textbf{\emph{5.90}}    &\textbf{\emph{5.9}}
                          &6.8&6.7    &6.5   &5.35   &5.80                    \\
$^{250}$Cm  &0.234   &0.242   &0.250&0.24   &0.26     &---
                          &&\textbf{\emph{5.39}}
                          &5.3    &5.9&6.5    &6.5   &4.79   &5.40            \\
$^{250}$Cf  &0.240   &0.246   &0.250&0.28   &0.26     &0.298
                          &&5.85    &6.5    &7.1&6.8    &6.7
                          &\textbf{\emph{5.70}}                    &5.60      \\
$^{252}$Cf  &0.237   &0.246   &0.251&0.25   &0.26     &0.304
                          &&\textbf{\emph{5.33}}
                          &5.8    &6.1&6.7    &6.2   &5.26   &5.30            \\
   \hline \hline
\end{tabular}
% \end{threeparttable}
\end{center}
\noindent
\\ [1mm]
\small
$^a$ The uncertainties are less than 0.015; see Ref.~\cite{Pritychenko2016}  for
                                                                      details.\\
$^b$ The bold italic denotes that this fission barrier, among these   theoretical
                                 values, is relatively close to experimental data.
\begin{multicols}{2}
%%%%%%%%%%%%%%%%%%%%%%%%%%%%%%%%%%%%%%%%%%%%%%%%%%%%%%%%%%%%%%%%%%%%%%%%%%%%%%%%
%1234567890123456789012345678901234567890123456789012345678901234567890123456789
\begin{center}
\includegraphics[width=7cm]{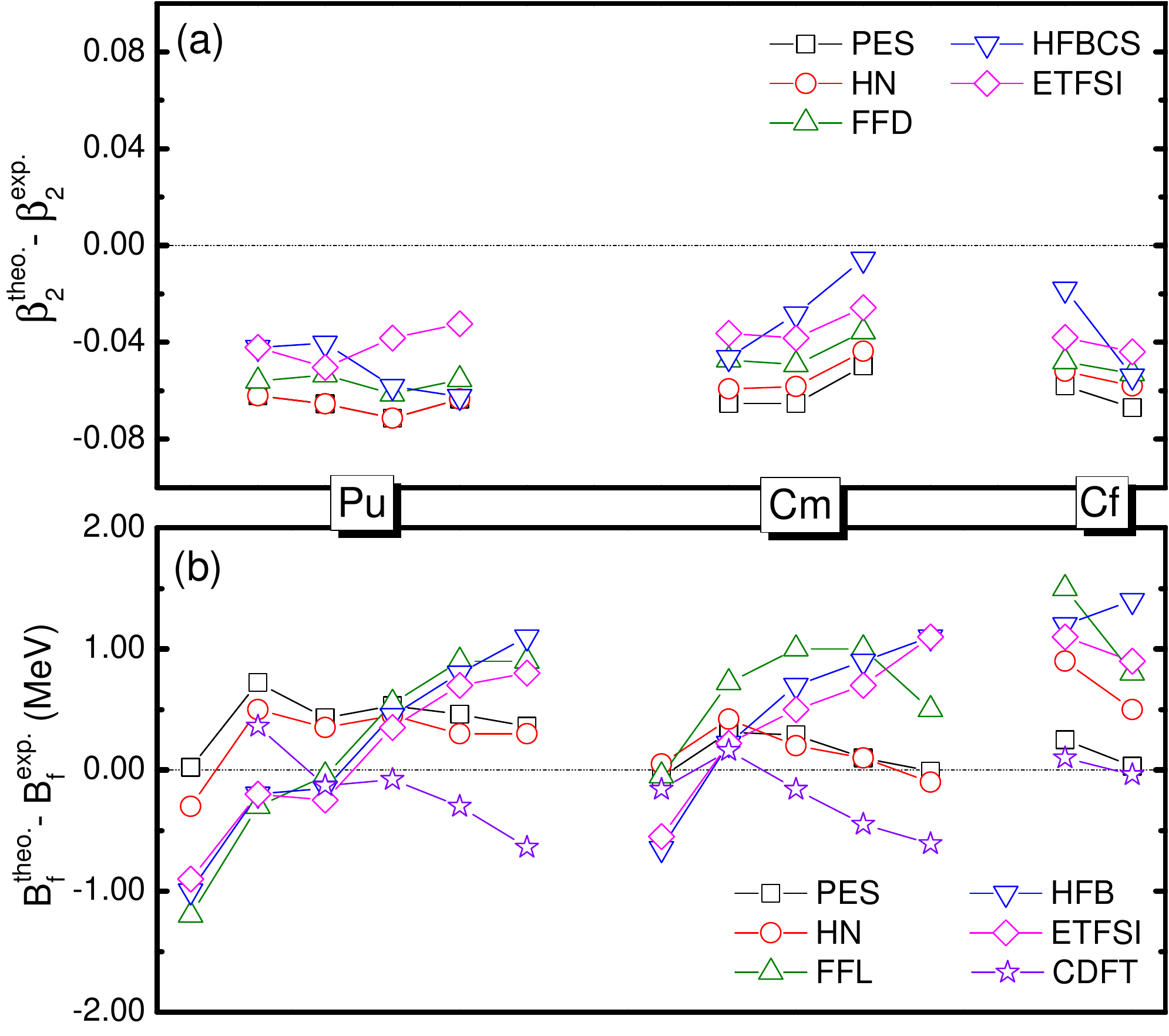}
\figcaption{
(a) The difference of quadrople deformation $\beta_2$ between theory         and
                            available data for transuranium Pu, Cm and Cf nuclei.
(b) Similar to (a) but for the difference of inner fission barrier $B_f$.
                                                                   \label{FIG.5}
}
\end{center}
%%%%%%%%%%%%%%%%%%%%%%%%%%%%%%%%%%%%%%%%%%%%%%%%%%%%%%%%%%%%%%%%%%%%%%%%%%%%%%%%
%description for Fig.5

\normalsize
Before continuing the investigation, we would like to further evaluate       the
         different theoretical results by comparing the differences between
                                                calculated and experimental data.
Figure~\ref{FIG.5}(a) and (b) show the differences between calculated       and
       experimental quadrople deformations $\beta_2$  and inner fission barriers
                                                             $B_f$, respectively.
It is seen that the calculated $\beta_2$ values are               systematically
                                      underestimated, especially in the present work.
Concerning this discrepancy, a corrected formula, e.g., for protons in the    WS
  case, $\beta_2^\rho \simeq 1.10\beta_2-0.03(\beta_2)^3$, has been suggested by
    Dudek $\emph{et al.}$~\cite{Dudek1984} by analyzing the relationship between
                     the WS potential parameters and the nucleonic distributions.
It is also pointed out that the vibration effect, e.g., the zero-point   motion,
       which would imply that the experiment-comparable deformations are not
       static values but rather are the most likely deformations calculated from the
    solutions of the collective motion, may be partly responsible for such shape
                                                inconsistency~\cite{Mazurek2015}.
From Fig.~\ref{FIG.5}(b), one can see that in most nuclei           theoretical
         calculations overestimate the fission barriers, especially in Cm and Cf
                                                                        isotopes.
It can be seen that the present PES calculation displays a relatively       high
                                                               descriptive power.
The rms (average) deviations of the present PES, HN, FFL, HFB, ETFSI and    CDFT
     results are 0.38 (0.26), 0.44 (0.28), 0.87 (0.48), 0.87 (0.45), 0.72 (0.34)
                                            and 0.32 ($-$0.16) MeV, respectively.
It seems that apart from the CDFT result, our calculation gives the         best
                                               description of experimental data.
However,  the CDFT result is somewhat far from  the         other
   theoretical calculations, and the very small average derivation may
     originate from the fluctuation of the calculated values above and below the
                zero line (the distribution of our results is mostly above zero).
In addition, our calculation gives a similar trend to the
                   HN calculation but has smaller rms and average deviations.
All these facts make us confident of the validity of our approach and of our  investigation of the impact of triaxiality  on the fission
                                                                         barrier.
%%%%%%%%%%%%%%%%%%%%%%%%%%%%%%%%%%%%%%%%%%%%%%%%%%%%%%%%%%%%%%%%%%%%%%%%%%%%%%%%
%description for Figs. 6 and 7

Taking the above facts into account, systematic multidimensional    PES
           calculations for 95 transuranium nuclei (including 49 actinide and 46
          superheavy members) have been carried out, focusing on the influence of triaxiality
                                           on the inner fission barrier.
Figures~\ref{FIG.6} and~\ref{FIG.7} show the calculated potential energy  curves
    with and without the triaxial deformation degree of freedom for these nuclei.
Note that the selected 95 nuclei ranging from Pu ($Z=94$) to Og        ($Z=118$)
                   isotopes, as seen in Figs.~\ref{FIG.6} and  \ref{FIG.7}, have
                                                        already been synthesized
         experimentally~\cite{NNDC,Oganessian2002,Oganessian2007,Oganessian2017}.
From these two figures, it is found that the ground states of these nuclei   are
      deformed and one can see the properties of the inner fission barriers
   including their heights, widths and the evolution with various nucleon numbers.
Obviously, for the actinide nuclei triaxiality has a considerable impact on  the
      inner fission barriers and makes the calculated barrier heights agree reasonably
                                                        with available data.
From the lower left to the upper right corner of Fig.~\ref{FIG.6},    the
        triaxiality decreasing the fission barrier widths gradually changes to
                                                              reducing the heights.
The fission path has been strongly modified by the triaxiality,    which
                         will lead to a significant reduction of the penetration
                                                probability~\cite{Sadhukhan2016}.
In the superheavy region, as shown in Fig.~\ref{FIG.7}, one can clearly see that
             the situation is more complicated than in the case of the actinides.
The barrier shape and the effect of triaxiality in the lighter Rf and     Sg
                    isotopes are similar to those in the heavier actinide nuclei.
However, in other nuclei the shapes of the fission barriers are rather different.
For instance, from the ``south-west'' to the ``north-east'' direction,       the
               fission barrier becomes wider and wider until it collapses in the
                                                            intermediate section.
Correspondingly, the energy curve, especially the triaxial one, becomes   softer
  and softer (i.e., there is no easily recognizable saddle point in the
      heavier Rf and Sg isotopes, Hs and Ds isotopes) until the bimodal barriers
                                     appear, e.g., in the FI, Lv and Og isotopes.
The bimodal barriers which appear are strongly affected   by
                                    the triaxiality, as seen in Fig.~\ref{FIG.7}.
%%%%%%%%%%%%%%%%%%%%%%%%%%%%%%%%%%%%%%%%%%%%%%%%%%%%%%%%%%%%%%%%%%%%%%%%%%%%%%%%
%1234567890123456789012345678901234567890123456789012345678901234567890123456789
\end{multicols}
\begin{center}
\includegraphics[width=15cm]{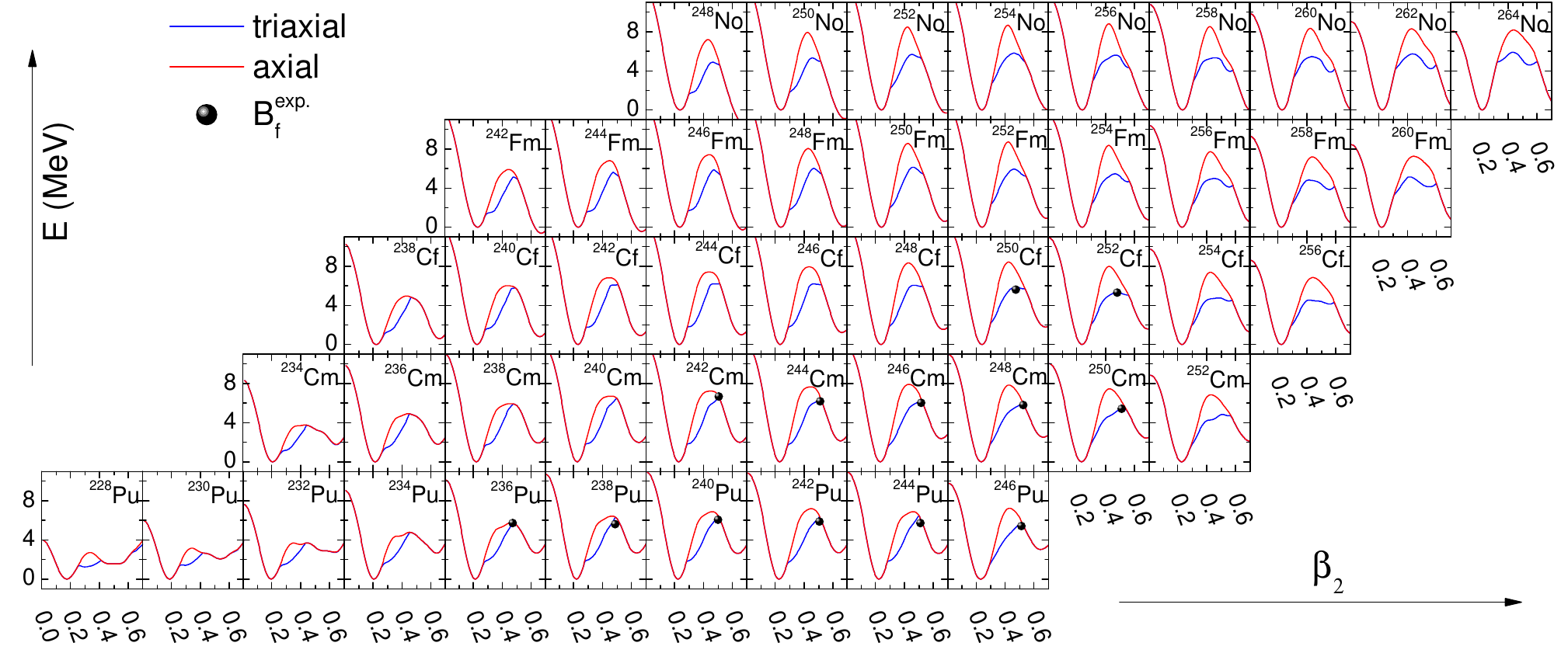}
\figcaption{
Similar to Fig.~\ref{FIG.1}(b), calculated potential energy curves of         49
                                even-even actinide nuclei with $94\leq Z\leq102$.
The solid black circles show the available data for the height of the inner       fission
                  barrier~\cite{IAEA1993,RIPL2009}, as seen in Table~\ref{table}.
                                                                   \label{FIG.6}
}
\end{center}
\begin{multicols}{2}
%%%%%%%%%%%%%%%%%%%%%%%%%%%%%%%%%%%%%%%%%%%%%%%%%%%%%%%%%%%%%%%%%%%%%%%%%%%%%%%%
%1234567890123456789012345678901234567890123456789012345678901234567890123456789
\end{multicols}
\begin{center}
\includegraphics[width=15cm]{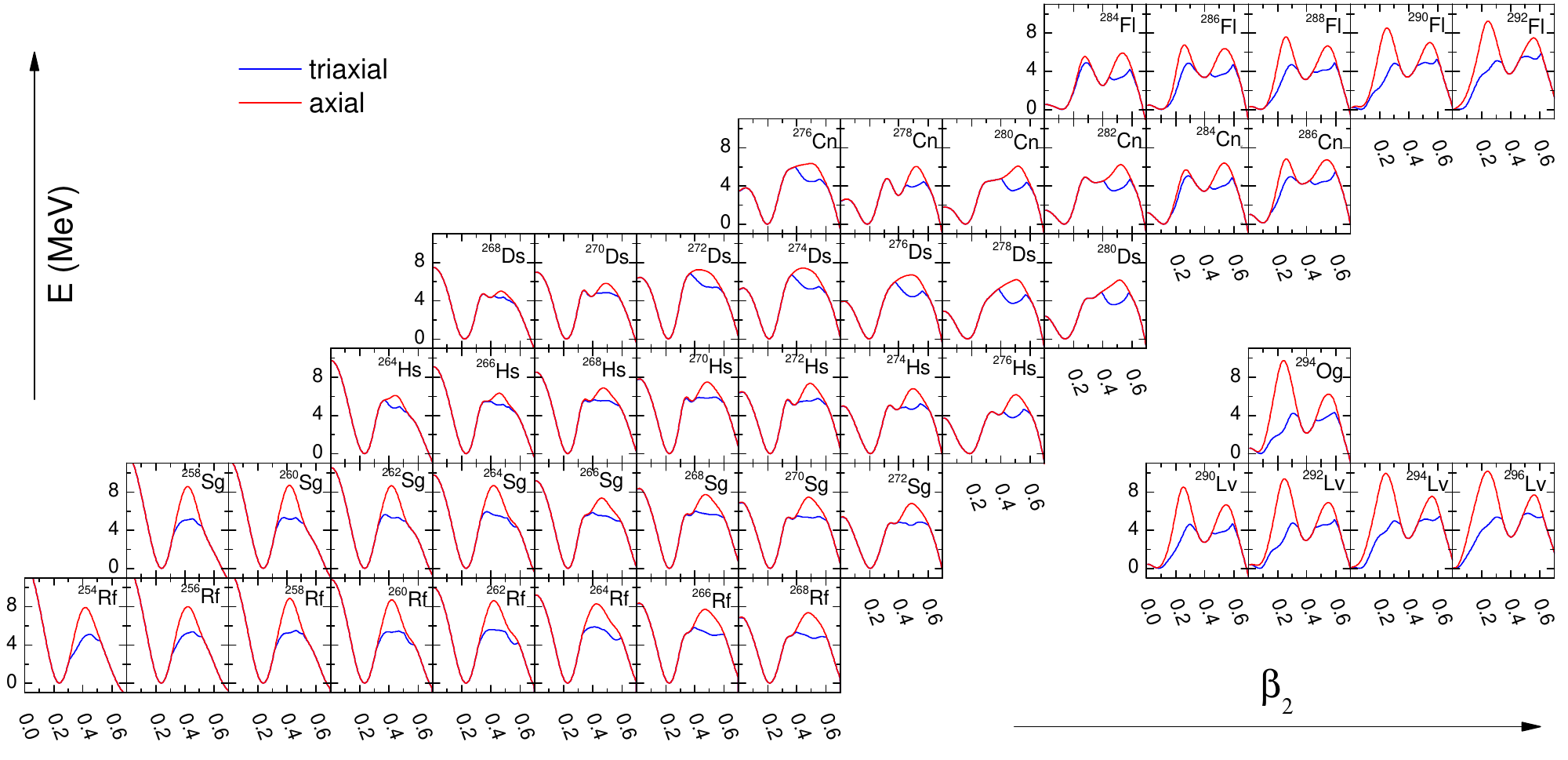}
\figcaption{
The same as in Fig.~\ref{FIG.6} but for 46 even-even superheavy nuclei      with
                                                              $104\leq Z\leq118$.
                                                                   \label{FIG.7}
}
\end{center}
\begin{multicols}{2}
%%%%%%%%%%%%%%%%%%%%%%%%%%%%%%%%%%%%%%%%%%%%%%%%%%%%%%%%%%%%%%%%%%%%%%%%%%%%%%%%
%123456789 123456789 123456789 123456789 123456789 123456789 123456789 123456789
%description for Figs. 8

\normalsize
In addition, in the MM calculation, besides a relatively large and    reasonable
   deformation space, the isospin dependence of the spin-orbit coupling strength
       and the nuclear surface diffuseness parameter has been found to be an
           important factor in the accurate description of nuclear ground-state
                                  properties, especially for the extreme isospin
                                                nuclei~\cite{WangN2014,ZhiQ2006}.
Similarly, aside from the fact that the inclusion of some critical   deformation
   degrees of freedom may greatly decrease the fission barrier, it was shown in our previous work  that the adjustment of the potential parameters (e.g., the strength
    of the spin-orbit potential, $\lambda$, and the nuclear surface diffuseness,
          $a$) can also affect the height of the fission barrier~\cite{YangJ2015}.
In the present transuranium region (in particular the superheavy region),  it is
     certainly of interest to examine to what extent the fission barrier will be
                   affected by a similar adjustment of the model parameters.
To search for high-isospin candidates far from decay stability, as a reference,
   an empirical formula $Z=A/(1.98+0.0155A^{2/3})$ for the $\beta$-stability line
   is used here, which denotes the location of the maximum of the binding energy
                                 per nucleon or the minimum of the $Q$-value for
                                       $\beta$-decay~\cite{Marmier1971,WuCL1996}.
With this in mind, we take the neutron-deficient superheavy $^{254}$Rf   nucleus
    as an example to investigate the effects of the spin-orbit coupling strength
         and the nuclear surface diffuseness on potential energy curves with and
                               without triaxiality, as shown in Fig.~\ref{FIG.8}.
The parameters ($a$, $\lambda$) are slightly modified between a narrow domain on
        the basis of the initial values (0.70, 36), namely the universal values,
               according to the isospin-dependent function relationship given in
                                                           Ref.~\cite{WangN2014}.
Of course, the increased combination (0.73, 38) of the          ($a$, $\lambda$)
      parameters for protons is accordingly expected in this proton-rich nucleus.
One can see from Fig.~\ref{FIG.8} that the fission barrier can be       slightly
    raised, indicating that the corresponding fission probability will increase
                                                                  to some extent.
A further test and improvement of model parameters, including the     macroscopic
                                            part, will be done in our future work.
%%%%%%%%%%%%%%%%%%%%%%%%%%%%%%%%%%%%%%%%%%%%%%%%%%%%%%%%%%%%%%%%%%%%%%%%%%%%%%%%
%1234567890123456789012345678901234567890123456789012345678901234567890123456789
\vspace{-18mm}
\begin{center}
\includegraphics[width=7cm]{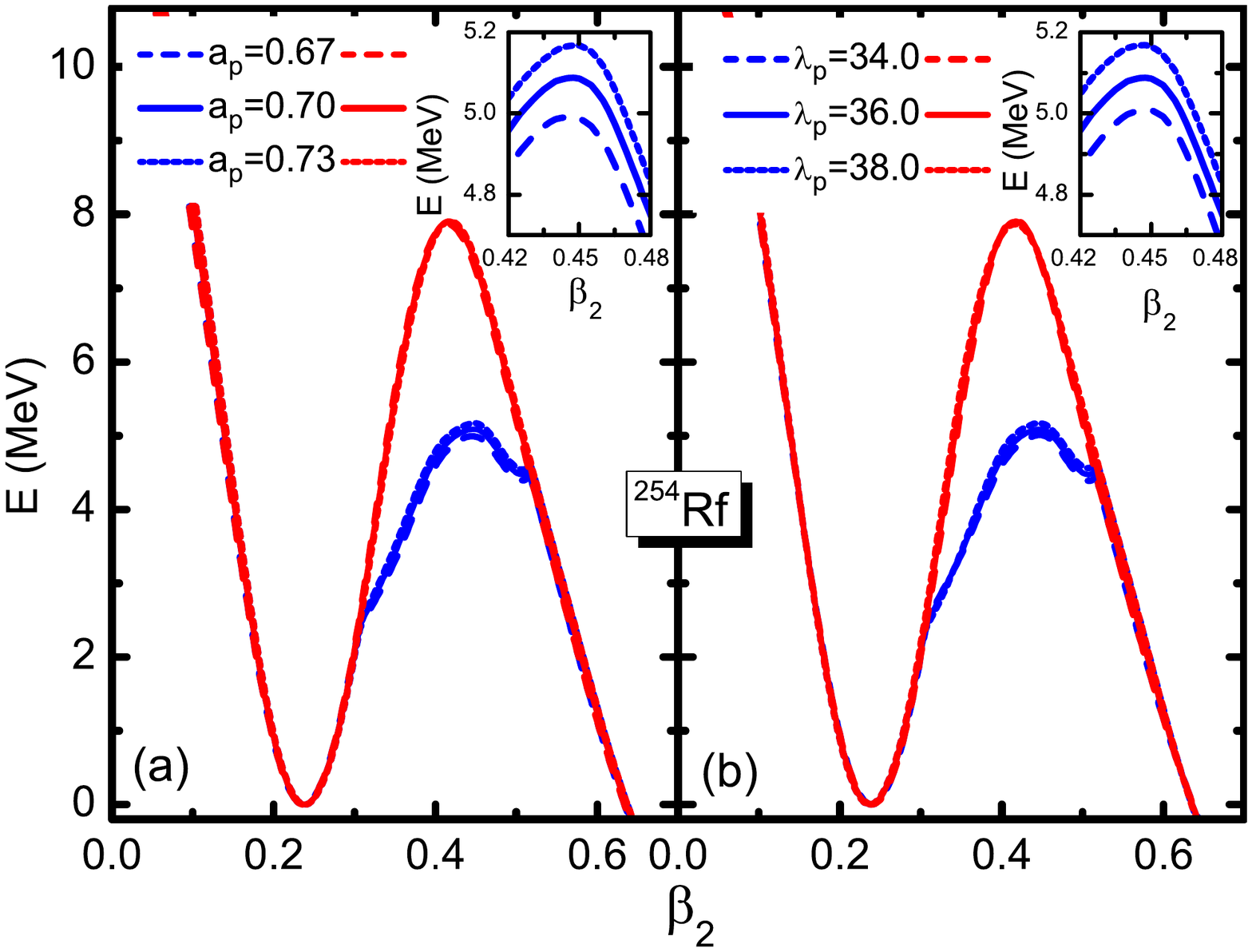}
\vspace{-20mm}
\figcaption{
Similar to Fig.~\ref{FIG.1}(b), but the energy curves with adjusted      surface
      diffuseness $a$ (a) and spin-orbit strength $\lambda$ (b) for the selected
                                            neutron-deficient nucleus $^{254}$Rf.
All other potential parameters are identical with those of      universal values.
The red (blue) lines denote the potential energy curves along the of       axial
                                                         (triaxial) fission path.
                                                                   \label{FIG.8}
}
\end{center}
%1234567890123456789012345678901234567890123456789012345678901234567890123456789
\section{Summary}\label{sec.IV}

In summary, we have systematically presented the axial and triaxial      fission
        barriers for 95 even-even transuranium nuclei ranging from $^{228}$Pu to
                                                                      $^{294}$Og.
The calculations were carried out using the pairing self-consistent  potential
             energy surface approach with universal Woods-Saxon parameter set in
    multidimensional ($\beta_2$, $\gamma$, $\beta_4$) deformation space with the
                                   inclusion of triaxial shape degree of freedom.
Our analysis shows that the main contribution of the fission barrier originates  from
     the microscopic energy, especially the shell correction, though the pairing
        correction and macroscopic energy sometimes play a rather important role.
Relative to the axially symmetric case, by allowing for triaxial    deformation,
             the height and/or width of the inner fission barrier may be reduced
   considerably in most nuclei (e.g., more than 4 MeV in $^{252}$Cf), which will
            lead to a significant increase of the penetration probability in the
                                                     spontaneous fission process.
A systematic comparison of the present results with experimentally        determined
     fission barriers shows a reasonable agreement (with rms deviation of 0.38
                                           MeV) in the available actinide nuclei.
Calculated energy curves with and without triaxial deformation in the   actinide
    and superheavy regions indicate the influence of triaxiality on not only the
     heights but also the widths of the fission barriers with increasing nucleon
                                                                          number.
Similar to mass calculations, it seems that to ensure the right fission barriers
            without unnecessary loss of CPU time, the selection of suitable
           deformation space is important, since missing any critical
            deformation degree of freedom may result in an overestimation of the
                                                             calculated barriers.
However, the underestimation of the theoretical calculations may be   attributed
                        to the choice of the potential parameters to some extent.
For instance, the adjustment of the strength of spin-orbit coupling and      the
             surface diffuseness will increase the height of the fission barrier.
Therefore, besides the macroscopic energy, the perfect combination of        the
   deformation space and potential parameter set will be a critical condition to
     guarantee good fission barriers in such phenomenological nuclear mean-field
                                                                    calculations.
This systematic investigation should be helpful to test and develop the model in
           future as well as to understand and predict the fission properties in
                                the even-even transuranium nuclei synthesized so far.
A similar formulation for dynamic fission barriers is currently under way     by
                          including the Coriolis effect in the model Hamiltonian.
%1234567890123456789012345678901234567890123456789012345678901234567890123456789

\end{multicols}
\vspace{-1mm} \centerline{\rule{80mm}{0.3pt}} \vspace{2mm}

\begin{multicols}{2}

\end{multicols}

\clearpage
\end{CJK*}
\end{document}